\documentclass[sigconf]{acmart}

\AtBeginDocument{%
  \providecommand\BibTeX{{%
    \normalfont B\kern-0.5em{\scshape i\kern-0.25em b}\kern-0.8em\TeX}}}

\setcopyright{acmlicensed}
\copyrightyear{2024}
\acmYear{2024}
\acmDOI{XXXXXXX.XXXXXXX}

\acmConference[KDD'24]{}{August 25--29,
  2024}{Barcelona, Spain}
\acmISBN{978-1-4503-XXXX-X/18/06}

\usepackage{makecell}
\usepackage{multirow}
\usepackage{algorithm}
\usepackage{algpseudocode}
\usepackage{tabularx}
\usepackage{graphicx}
\usepackage{enumitem}

\newcounter{savealgorithm}
\newenvironment{subalgorithms}
 {%
  \stepcounter{algorithm}%
  \edef\currentthealgorithm{\thealgorithm}%
  \setcounter{savealgorithm}{\value{algorithm}}%
  \setcounter{algorithm}{0}%
  \renewcommand{\thealgorithm}{\currentthealgorithm\alph{algorithm}}%
 }
 {%
  \setcounter{algorithm}{\value{savealgorithm}}%
 }

\makeatother

\algnewcommand\algorithmicforeach{\textbf{for each}}
\algdef{S}[FOR]{ForEach}[1]{\algorithmicforeach\ #1\ \algorithmicdo}

\algdef{SE}[REPEATN]{RepeatN}{End}[1]{\algorithmicrepeat\ #1 \textbf{times}}{\algorithmicend}

\begin{document}

\title{LiMAML: Personalization of Deep Recommender Models via Meta Learning}

\author{Ruofan Wang, Prakruthi Prabhakar, Gaurav Srivastava, Tianqi Wang, Zeinab S. Jalali, Varun Bharill, Yunbo Ouyang, Aastha Nigam, Divya Venugopalan, Aman Gupta, Fedor Borisyuk, Sathiya Keerthi, Ajith Muralidharan$^{*}$}
    
\affiliation{%
  \institution{LinkedIn Inc}
  \country{USA}
}



\renewcommand{\shortauthors}{Wang and Prabhakar, et al.}

\begin{abstract}
In the realm of recommender systems, the ubiquitous adoption of deep neural networks has emerged as a dominant paradigm for modeling diverse business objectives. As user bases continue to expand, the necessity of personalization and frequent model updates have assumed paramount significance to ensure the delivery of relevant and refreshed experiences to a diverse array of members. In this work, we introduce an innovative meta-learning solution tailored to the personalization of models for individual members and other entities, coupled with the frequent updates based on the latest user interaction signals. Specifically, we leverage the Model-Agnostic Meta Learning (MAML) algorithm to adapt per-task sub-networks using recent user interaction data. Given the near infeasibility of productionizing original MAML-based models in online recommendation systems, we propose an efficient strategy to operationalize meta-learned sub-networks in production, which involves transforming them into fixed-sized vectors, termed meta embeddings, thereby enabling the seamless deployment of models with hundreds of billions of parameters for online serving. Through extensive experimentation on production data drawn from various applications at LinkedIn, we demonstrate that the proposed solution consistently outperforms the baseline models of those applications, including strong baselines such as using wide-and-deep ID based personalization approach. Our approach has enabled the deployment of a range of highly personalized AI models across diverse LinkedIn applications, leading to substantial improvements in business metrics as well as refreshed experience for our members.
\end{abstract}



\maketitle

\def\thefootnote{*}\footnotetext{Work done while at LinkedIn.}
\def\thefootnote{\arabic{footnote}}

\section{Introduction}\label{sec:intro}

With the advent of deep learning, neural-network based recommendation models have become very popular for modeling a variety of objectives at large internet companies, including click-through rate (CTR) prediction, invite prediction, and visit prediction. It is important for these models to understand the diverse and evolving individual needs of millions of members and provide refreshed item recommendations everyday.

An array of approaches have been proposed to improve the performance of such networks for better personalization. The first set of approaches propose architectural improvements to learn sophisticated feature interactions ~\cite{cheng2016wide, guo2017deepfm, wang2021dcn}. Deeper models with intricate architectures based on factorization machines have achieved state of the art performance on several recommender system prediction tasks. However, they are still global models served across all members, entities and items with limited personalization. The second set of approaches leverage the power of embeddings to handle sparse categorical feature inputs per user or per entity (e.g. per advertiser, per job, per industry segment, etc.) ~\cite{qu2016product, naumov2019deep, zhao2018learning}. Embedding based approaches provide some degree of personalization for each user or entity. But such approaches can only learn reliable embeddings when there is a significant amount of data per user, limiting their personalization scope to frequent members on the platform. In other words, such models are data inefficient during learning.

In general, the definition of personalization varies across different objectives and applications. As an example, for an ad - Click Through Rate (CTR) prediction task, we may want to build personalized models for each advertiser ID. For a general user-item CTR prediction task, we may want to build personalized models per user. For a model that predicts whether or not a user would apply for a job, a per-user and job-to-industry segment level personalization might make more sense. In the meta learning framework\footnote{See Section \ref{sec:preliminary} for preliminaries on meta learning.}, this would entail designing different task definitions for different use cases.

With meta learning, the goal is to quickly and effectively learn a new task from a small number of data samples using a model that is learnt on a large number of different tasks. ~\cite{finn2017} proposes a model-agnostic meta-learning algorithm (MAML) that modifies the optimization approach so that it can be directly applied to any deep learning model that is trained with a gradient descent procedure. Only recently, MAML based approaches have started gaining popularity in recommendation systems for handling cold start scenarios of CTR prediction models ~\cite{lee2019melu, pan2019warm, vartak2017meta}.  These approaches try to learn personalized user preferences by meta learning ID embedding vectors for cold start items or meta learning final few layers of the Multi-Layer Perceptron (MLP) based on few user interaction samples. 

In this paper, we extend all of these ideas to provide a general architecture for personalizing deep recommender models via meta learning. In our approach, we leverage MAML ~\cite{finn2017} to meta learn a part of the network at a per-task level, where tasks could be members or entities or a combination of both. The rest of the network is shared across all tasks and is kept identical to the architecture for that application. As opposed to the original MAML which is nearly infeasible to be productionized online in large scale recommendation systems, this approach provides an easy way to deploy personalized models with hundreds of billions of parameters in production by converting the output of the meta-learnt sub-network into meta embeddings, and providing these embeddings as regular features while serving the deployed neural network. By doing meta-finetuning frequently, we present a way to keep the models refreshed and personalized to the most recent user interactions on the platform. We demonstrate across several real-world use cases at LinkedIn that the proposed approach (termed LiMAML) beats the baseline models across all use cases. Along with business impact, we dive deep into how this approach truly uplifts model performance even for infrequent members with limited data, thereby presenting a compelling case for adopting a meta learnt personalization strategy for all recommender system models.

We go over related work in Section ~\ref{sec:related_work} and provide preliminaries on meta learning in Section ~\ref{sec:preliminary}. In Section ~\ref{sec:methodology_new}, we introduce our proposed methodology and in Section ~\ref{sec:serving}, we share some insights around LiMAML productionization. Offline experiments and online A/B testing results are presented in Sections ~\ref{sec:experiments} and ~\ref{sec:online} respectively.

\section{Related Work}\label{sec:related_work}
Deep recommender models usually rely on deep feature interactions and large ID embedding tables to enable personalization. Feature interaction modules such as Wide-and-Deep Network~\cite{cheng2016wide}, DeepFM~\cite{guo2017deepfm} and DCN v2~\cite{wang2021dcn} effectively fuse user features and entity features to improve the representation power. Another stream of approaches attempt to learn a personalized embedding per user or entity by introducing large cardinality embedding tables ~\cite{peng2021learning, qu2016product, naumov2019deep, zhao2018learning}. Although the aforementioned recommendation approaches can provide user-level or entity-level personalization, in industrial-scale recommender applications with billions of members and entities, these approaches require large models to incorporate personalized information and large data to train such models, making it less feasible in some large scale online recommendation systems due to limited data availability.

Meta learning, on the contrary, focuses on training models to adapt to every user or entity quickly when only a few samples are provided. Most literature on meta learning focuses on vision ~\cite{finn2017, rajeswaran2019meta} and language domain ~\cite{hu2023meta}, whereas applications of recommender systems domain have unique challenges since the number of tasks are much larger than vision and language domain. Meta learning can be roughly classified into 3 categories: model-based ~\cite{santoro2016meta}, metric-based ~\cite{snell2017prototypical} and optimization-based ~\cite{finn2017}. Multiple works have applied metric-based and model-based meta learning for recommender systems ~\cite{luo2020metaselector}, applied meta learning for scenario specific task definition ~\cite{du2019sequential} and applied meta learning in the online setting ~\cite{kim2022meta, peng2021learning}. These recommendation approaches focus on small-scale, non-industrial benchmark datasets, such as movie recommendation ~\cite{harper2015movielens}, book recommendation ~\cite{ziegler2005improving}, ads prediction ~\cite{kddcup2012-track2}.

Model Agnostic Meta Learning (MAML)~\cite{finn2017} enables large-scale task adaptation through bi-level optimization setup. MAML based optimization has already been applied for recommender systems ~\cite{wang2022, lee2019melu, neupane2022dynamic, hu2021hybrid, kim2023meta, vartak2017meta}. For example, MAML-based recommender systems have been utilized for ads CTR prediction problems ~\cite{pan2019warm, pang2022pnmta}. Meta-Learned User Preference Estimator (MeLU) ~\cite{lee2019melu} performs MAML based task adaptation on the MLP layers of the model. However, productionizing MeLU type architectures is challenging because the task adaptation layers are the last layers in the model architecture, which necessitates the storing of the personalization weights of all the members. Instead, we perform task adaptation on a sub-network, which includes only the first few layers of the network. This sub-network outputs a fixed size meta learned embedding vector during the inference for each user, which significantly facilitates model serving. 

A few works have applied Meta learning based solutions for recommender systems on production scale datasets ~\cite{yu2021personalized, ouyang2021learning}. However, these papers do not provide meta learning specific productionization strategies for industrial scale recommender systems. G-Meta ~\cite{xiao2023g} discussed speeding up MAML-based recommender systems by using GPU parallelism strategies, but other optimization strategies can be utilized together with parallelism to further accelerate meta learning model training.

\section{Preliminaries}\label{sec:preliminary}

The central idea of meta learning, also referred to as "learning-to-learn", is to learn a model \( \theta\) across a variety of learning tasks, such that it is capable of adapting to any new task quickly with only a small number of data points. There are broadly three types of meta learning: metric based, model based and optimization based. In this paper, we adopt optimization based meta learning from the reference paper ~\cite{finn2017}. In this approach called Model-Agnostic Meta Learning (MAML), we learn the model parameters \( \theta\) such that the model has maximal generalization performance on a new task after the parameters have been updated through one or more gradient steps to a personalized \( \theta_i \) starting from \( \theta\).

Formally, let us consider a prediction function \(f_\theta: x \rightarrow y\) which maps observations denoted by \(x\) to outputs denoted by \(y\). \(f\) can be any neural network based function approximator with parameters \(\theta\). Then we define each \textbf{task} as \(T_i = \{(x_1, y_1), (x_2, y_2), \ldots)\}\) where \(x_j, y_j\) are i.i.d. samples from a specific task \(T_i\). The loss \(\mathcal{L}_{T_i}(f_{\theta_i})\) provides task-specific feedback based on the problem type using the task-specific model weights \( \theta_i \). For a binary classification problem, it can be cross entropy loss, and for a regression problem, it can be mean-squared error (MSE) loss. We want the meta learnt model to be performing well on a distribution of learning tasks \(p(\mathcal{T})\). The entire dataset is therefore constructed as a set of tasks \(\{ T_1, T_2, \ldots, T_N \}\) where \(N\) is the number of tasks in total and each \(T_i \sim p(\mathcal{T})\) refers to a single task with all data points under task \(i\). Each task level data \(T_i\) is further split into two parts: \textbf{support} set and \textbf{query} set. The support set is utilized for task-level personalization and query set is utilized for maximizing generalization performance across tasks. This will become clearer as we describe the steps of MAML given in Algorithm~\ref{alg:original_maml}\cite{finn2017}.

The MAML algorithm mainly involves two stages: task adaptation phase (Lines 4 - 10 of Algorithm~\ref{alg:original_maml}), henceforth called the inner loop and meta-optimization phase (Line 11 of Algorithm~\ref{alg:original_maml}), henceforth called the outer loop. The goal of the inner loop is to learn task-level personalization. It does so by minimizing the loss on each task's support set data by performing gradient updates \(n\) times to obtain a set of personalized (fine-tuned) model weights \( \theta_i\) per task. Note that at a task-level, the learning process presents an over-parameterized problem, with multiple solutions for \(\theta_i\) that can minimize the loss on the support set. However, this algorithm restricts the solution space by bootstrapping from \(\theta\) as the starting point to learn \(\theta_i\), creating a strong dependence of \(\theta_i\) on \(\theta\). The step size \(\alpha\) is the task learning rate. \(\nabla_\theta\) refers to the gradient w.r.t. \(\theta\). The inner loop is also sometimes referred to as task-level fine-tuning as we fine tune \( \theta\) to learn personalized model parameters \( \theta_i\) for each task using a few samples from it.

The goal of the outer loop is to update the model parameters \(\theta\) such that it can maximize the generalization performance on a wide variety of tasks. This is achieved by doing gradient update  ~\cite{finn2017} of \(\theta\) using the losses computed on the query sets of each task from the per-task model parameters \(\theta_i\). Note that this gradient update is done using all tasks since \(\theta\) is shared across all tasks. \(\beta\) is the learning rate used in the meta optimization step, also known as global learning rate, and is usually different from the task learning rate \(\alpha\). The minimization of the losses of different \(\theta_i\)s computed on the query sets represents the maximization of generalization performance across tasks. The outer loop update involves a gradient through a gradient computation, which requires Hessian-vector products computation.

In essence, we learn the model parameters \( \theta\) such that the model has maximal generalization performance on any new task after the parameters for that task are bootstrapped from \(\theta\) and updated through one or more task-level gradient steps to a personalized \(\theta_i\).

\begin{algorithm}[htbp!]
\small
\caption{Original MAML Algorithm \cite{finn2017}}
\label{alg:original_maml}
\begin{algorithmic}[1]
\Require \( p(\mathcal{T}) \): distribution over tasks
\Require \( \alpha, \beta\): step size/learning rate hyperparameters of inner and outer loop
\Require \(n\): number of times to repeat the inner loop gradient updates
\State randomly initialize \( \theta \)
\While{not done}
    \State Sample batch of tasks \( T_i \sim p(\mathcal{T}) \)
    \ForAll{\( T_i \)}
        \State \( \theta_i \gets \theta\)
        \RepeatN{$n$}
        \State Evaluate \( \nabla_{\theta_i} \mathcal{L}_{T_i}(f_{\theta_i}) \) with support set
        \State \( \theta_i \gets \theta_i - \alpha \nabla_{\theta_i} \mathcal{L}_{T_i}(f_{\theta_i}) \)
        \End
    \EndFor
    \State Update \( \theta \gets \theta - \beta \nabla_{\theta} \sum_{T_i \sim p(\mathcal{T})} \mathcal{L}_{T_i}(f_{\theta_i}) \) with query set
\EndWhile
\end{algorithmic}
\end{algorithm}

\section{Methodology}\label{sec:methodology_new}

In this section, we start with discussing how MAML can be designed for applications in recommender systems, specifically choosing the right task definitions and loss functions. We will then present the difficulties of deploying MAML based solutions as is in production. After that, we introduce our proposed method: \textbf{LiMAML}, which preserves the benefits from meta learning while also ensuring a scalable and production-friendly approach to personalization.

\subsection{Task definition in MAML}

Defining the task is one of the critical first steps in formulating the problem via meta learning. The goal of personalizing a model via meta learning is to effectively and quickly learn to produce the fine-tuned network weights for each new task. If the learning objective is to predict CTR on an item from a user, each user or user segment could be a natural choice of task definition for this problem. In general, the input to the neural network in recommender systems usually consists of a set of one or more entities and their corresponding features. Examples of entities include job ID, advertiser ID, viewer ID, etc. One or more combination of entities or their segmentation would be a good choice in designing the task for meta learning. If we treat each viewer ID as a task, then \(T_i\) includes all data points for a particular viewer and the outcome of meta learning would be to produce per-viewer personalized networks based on most recent viewer interaction data.

\subsection{Loss functions for a task}
The next step would be to define the loss function for the learning problem. In recommender systems, the prediction function \(f_\theta: x \rightarrow y\) is usually a binary classifier probability predicting a user response given a set of features. \(x\) is a collective feature set including several dimensions such as user, item, context, etc. \(y\) represents the user response, for example, whether a user clicked on a recommended item. \(f\) can be of any form. In this paper, we assume \(f\) to be a neural network based function approximator. Given \(y\) is a binary signal in general, \(\mathcal{L}\) can be cross-entropy loss as defined below. Please note that in MAML, the losses are defined at a task level.

\begin{equation}
\mathcal{L}_{T_i}(f) = \sum_{j} \left[ y_j \log(f(x_j)) + (1 - y_j) \log(1 - f(x_j)) \right]
\end{equation}

\subsection{LiMAML}

With the MAML algorithm, by treating each entity as a task, we can build a personalized model where each entity has its own set of values for the fine-tuned model parameters. In order to deploy such a model in practice, we would either have to store the personalized per-entity network weights and serve them during inference or would have to do the fine-tuning during inference in real time. Both of these solutions are infeasible to be deployed into production in online recommendation systems mainly from two aspects.

\textbf{Storage}: If we treat each user or each entity as a task, the number of tasks can be extremely large, running in the order of millions or more. For example, there are over 1 billion members of LinkedIn and it is extremely expensive to store 1 billion \(\theta_i\) per user. Hence, storing the full \( \theta_i \forall i \) would be infeasible from a storage perspective.

\textbf{Latency}: Recommendations are made online in near real time. Performing task-level fine-tuning during an inference call could cause significant inference latency with high computation cost. Hence, doing real time fine-tuning would be infeasible from a compute perspective.

\begin{figure}[htbp!]
    \centering
    \includegraphics[width=0.3\textwidth]{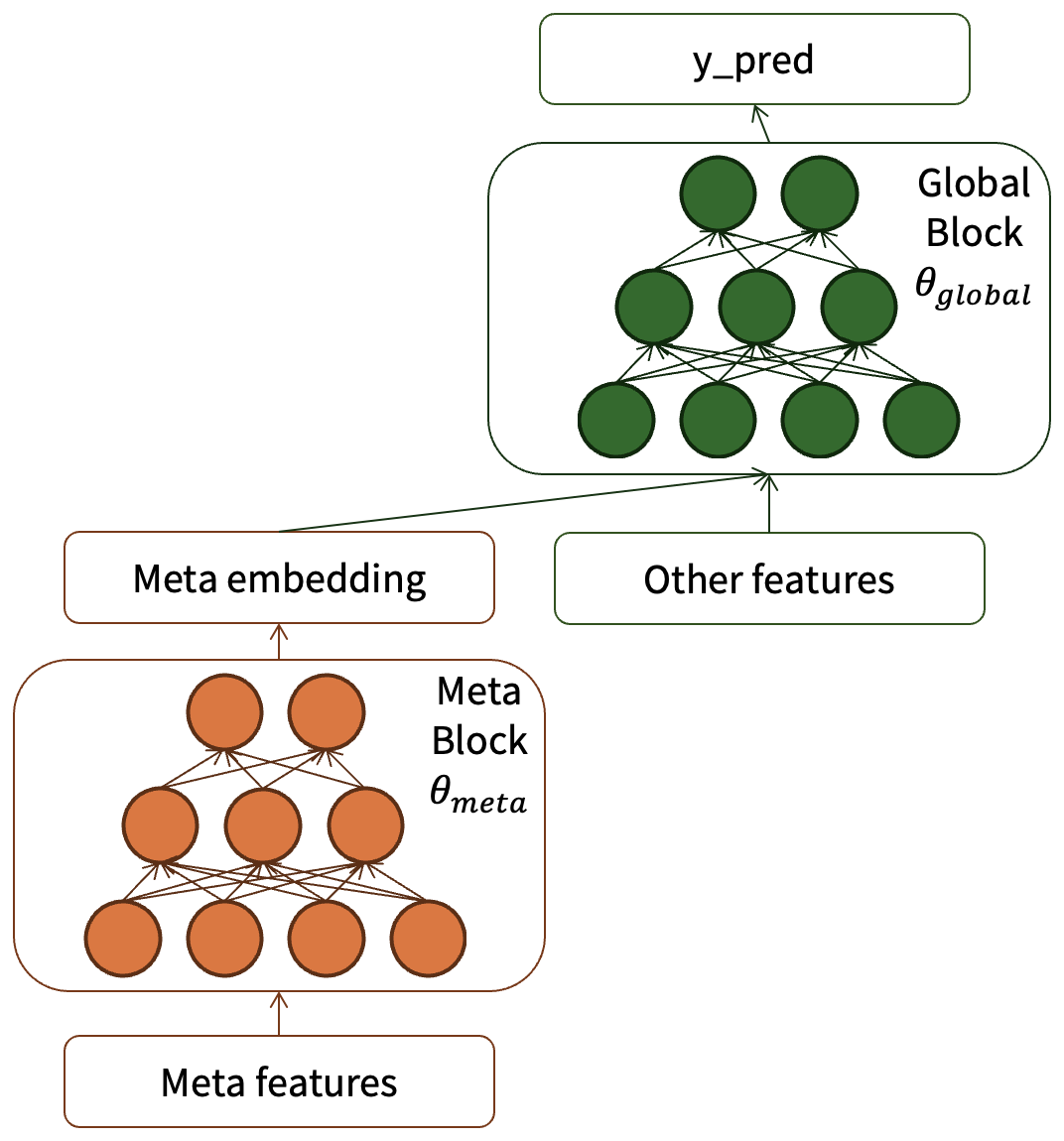}
    \caption{LiMAML Model Structure}
    \label{fig:partial_maml_model}
    \vspace{-1.0em}
\end{figure}

These two considerations inspired us to come up with an effective variant of the original MAML for LinkedIn recommender applications, called \textbf{LiMAML}. It still possesses the capability of personalization, offering well-tailored and frequently refreshed models per entity, but more importantly, provides a much easier framework to productionize and serve at a larger scale. Algorithms \ref{alg:partial_maml_training} and \ref{alg:partial_maml_inference} provide an overview of the algorithm. Figure \ref{fig:partial_maml_model} provides the network structure of LiMAML.

\begin{subalgorithms}
\begin{algorithm}[htbp!]
\small
\caption{LiMAML: \textbf{Training}}
\label{alg:partial_maml_training}
\begin{algorithmic}[1]
\Require \( p(\mathcal{T}) \): distribution over tasks
\Require \( \alpha, \beta \): step size/learning rate hyperparameters of inner and outer loop
\Require \(n\): number of times to repeat the inner loop gradient updates
\State randomly initialize \( \theta_{\text{meta}} \), \( \theta_{\text{global}} \)
\While{not done}
    \State Sample batch of tasks \( T_i \sim p(T) \)
    \ForAll{\( T_i \)}
        \State \( \theta_{\text{meta}_i} \gets \theta_{\text{meta}}\)
        \RepeatN{$n$}
            \State Evaluate \( \nabla_{\theta_{\text{meta}_i}} \mathcal{L}_{T_i}(f_{\theta_{\text{meta}_i}}) \) with support set 
            \State \( \theta_{\text{meta}_i} \leftarrow \theta_{\text{meta}_i} - \alpha \nabla_{\theta_{\text{meta}_i}} \mathcal{L}_{T_i}(f_{\theta_{\text{meta}_i}}) \)
        \End
    \EndFor
    \State Update \( \theta_{\text{meta}}\) and \( \theta_{\text{global}}\) with query set:
    \State \( \theta_{\text{meta}} \leftarrow \theta_{\text{meta}} - \beta \nabla_{\theta_{\text{meta}}} \sum_{i=1}^{N} \mathcal{L}_{T_i}(f_{\theta_{\text{meta}_i}}) \) 
    \State \( \theta_{\text{global}} \leftarrow \theta_{\text{global}} - \beta \nabla_{\theta_{\text{global}}} \sum_{i=1}^{N} \mathcal{L}_{T_i}(f_{\theta_{\text{global}}}) \) 
\EndWhile
\end{algorithmic}
\end{algorithm}
\vspace{-0.2in}
\begin{algorithm}[htbp!]
\small
\caption{LiMAML: \textbf{Meta Embedding Generation}}
\label{alg:partial_maml_inference}
\begin{algorithmic}[1]
\Require \(k\): number of times to repeat the fine-tuning gradient updates
\ForEach {$T_i \in \mathcal T $}
    \State \( \theta_{\text{meta}_i} \gets \theta_{\text{meta}}\)
    \RepeatN{$k$}
        \State Evaluate \( \nabla_{\theta_{\text{meta}_i}} \mathcal{L}_{T_i}(f_{\theta_{\text{meta}_i}}) \) with recent samples of \(T_i\)
        \State \( \theta_{\text{meta}_i} \leftarrow \theta_{\text{meta}_i} - \alpha \nabla_{\theta_{\text{meta}_i}} \mathcal{L}_{T_i}(f_{\theta_{\text{meta}_i}}) \)
    \End
    \State Score the most recent sample \(x_i\) using \( \theta_{\text{meta}_i}\) to obtain the output of the meta block (meta embedding) \(E_i\)
\EndFor
\end{algorithmic}
\end{algorithm}
\end{subalgorithms}

\subsubsection{Network Structure}
\label{sec:network-structure}
In this approach, we divide the network into two blocks - \textbf{Meta Block} and \textbf{Global Block}. Meta block, whose parameters are denoted by \( \theta_{\text{meta}} \), is the sub-network that we will meta learn using MAML approach. Hence, for every task, we will start with \( \theta_{\text{meta}} \), and fine-tune it to produce personalized sub-networks \( \theta_{\text{meta}_1} \), \( \theta_{\text{meta}_2} \), \ldots, \( \theta_{\text{meta}_N} \) for \(N\) tasks. Global block, whose parameters are denoted by \( \theta_{\text{global}} \), is the sub-network that is shared across all tasks and is usually equivalent in architecture to the network structure currently deployed for an application. We will split the input features for each task into two categories. \textbf{Meta Features} are the input to the meta block, which usually contains features specific to the entity which forms the task definition. For example, if we are meta learning the sub-network per user, meta features are the user features for each item in the task. \textbf{Other Features} refer to all other features input to the global model. In our applications of LiMAML, meta features can also be fed directly into the global block as well. Global block additionally takes in the output of the meta block, referred to as \textbf{Meta Embedding}, as additional inputs along with other features. In our experiments, we simply concatenated meta embeddings with other features and provided that as input to the Global block. In essence, the meta block provides the personalized embeddings per task as additional input signals to the global block.

\subsubsection{Training}
Algorithm \ref{alg:partial_maml_training} provides the steps to train LiMAML. In the inner loop (Lines 4 - 10), we only update the meta block parameters \( \theta_{\text{meta}_i}\) for each task \( T_i \) using the \textit{support} set data from each task. In the outer loop (Lines 11 - 13), both meta block parameters \( \theta_{\text{meta}}\) and global block parameters \( \theta_{\text{global}}\) are updated with the \textit{query} set of training data. For the meta block parameter update, the loss for the gradient is computed using each task's fine-tuned model parameters \(\theta_{\text{meta}_i}\). For the global block parameter update, the loss for the gradient is computed using the model parameters \(\theta_{\text{global}}\). By the end of training, we have learned a set of model parameters \(\langle \theta_{\text{meta}}, \theta_{\text{global}} \rangle\) for both blocks. \(\theta_{\text{meta}}\) is obtained by training against a variety of tasks and therefore capable of adapting quickly to any new or old task given just a few data points.

\subsubsection{Meta Embedding Generation}
The central idea of LiMAML is to decouple serving of meta block and global block in production. \textit{Meta block} is served offline, whereas \textit{Global block} is served online during inference. In order to do this, we create a separate flow called \textbf{Meta Embedding Generation}, which is run on a regular cadence (say, once per day) to do fine-tuning of the meta block to output \textit{meta embeddings}. Algorithm \ref{alg:partial_maml_inference} provides the set of steps for updating \(\theta_{\text{meta}}\) frequently via meta fine-tuning and producing embeddings for online serving of the global block. In lines 3-6, we use recent samples of each task to update \(\theta_{\text{meta}}\) and obtain \(\theta_{\text{meta}_i}\) for that task by taking \(k\) gradient descent steps. Note that the number of gradient steps \(k\) can be different from the number of inner loop gradient steps \(n\) taken during training. Every time we run the meta embedding generation flow, we bootstrap the meta block parameters with \(\theta_{\text{meta}}\) as indicated in line 2 of the algorithm. After getting \(\theta_{\text{meta}_i}\) for each task, we then immediately score the meta block with \(\theta_{\text{meta}_i}\) as the model parameters using the most recent sample \(x_i\) for that task as shown in line 7. We then store the output of the meta block as meta embedding \(E_i\) for that task. Note that the input to the meta block only contains entity-specific features and do not contain any other item specific features for a task. Hence, scoring the personalized meta block with the most recent sample \(x_i\) would correspond to scoring with the latest entity specific features and obtaining an embedding for that entity. In our experiments, we also tried variants of this approach such as mean pooling of meta embeddings across multiple samples of a task and found little to no difference in the performance of our model. Instead of persisting all the updated task-level model parameters, we only store the output of meta block, which is a fix-sized vector, so called \textit{meta embedding}. These embedding vectors will be persisted and stored in a feature store for retrieval during online inference. This is an extremely important change as it reduces the required storage from a set of model weights per task (quadrillions of parameters) to an embedding vector per task (tens of billions of parameters). \textit{Global Block} is served online as per the usual deployment and inference process. When a new scoring request comes in, we retrieve all features as well as the latest version of \textit{meta embeddings} from the feature store, and score them with the global block \(\theta_{\text{global}}\).
 
We highlight the differences between the original MAML and our approach LiMAML in Table \ref{tab:orimaml_partialmaml}. In summary, the network is decoupled into two blocks: Meta Block and Global Block. With LiMAML, some of the key advantages include
\begin{itemize}[leftmargin=*]
\item We are able to preserve the personalization capability, while enabling easier deployment of large scale models.
\item Given that meta block is served offline using a sequence of samples per task, we can use any simple to complex architecture for personalization such as ID embedding layer, dense MLP or a transformer for the meta block.
\item We can easily refresh a part of the model at a regular cadence and adapt it to the recent user interactions occurring on the platform.
\end{itemize}

\begin{table}[htbp!]
\small
\centering
\begin{tabular}{p{1.1cm}|p{3.1cm}|p{3.1cm}}
\toprule
& \textbf{Original MAML} & \textbf{LiMAML}  \\
\hline
\textbf{Network} & not decoupled & decoupled into Meta Block and Global Block \\ 
\hline
\textbf{Training} & entire network is meta learned & only Meta Block is meta learned \\
\hline
\textbf{Serving} & entire network needs to be served online & hybrid serving, meta block is served offline, global block is served online \\
\hline
\textbf{Storage} & required to store all model parameters for all tasks & store an embedding vector per task \\
\hline
\textbf{Latency} & huge inference latency if fine-tuning happens online & no inference latency overhead as fine-tuning happens offline \\
\bottomrule
\end{tabular}
\caption{Original MAML v.s. LiMAML.}
\label{tab:orimaml_partialmaml}
\vspace{-2.0em}
\end{table}

\section{Production Insights}\label{sec:serving}

Serving refreshed meta embeddings via frequent fine-tuning brings in additional complexities to the data and model pipelines in production, especially when the number of tasks runs to the order of hundreds of millions. In this section, we present some insights on how we designed our pipelines and some lessons we learnt while deploying LiMAML to production.

Every day, there are two pipelines running in production: a data pipeline and an inference pipeline. As members are interacting with the platform, we first needed a daily data processing pipeline that collects these interactions and creates labeled data by joining with feature datasets. We follow this with an inference pipeline that collects the most recent \(X\) days of data, groups it on a task level, and performs meta fine-tuning to produce \textit{meta embeddings} for all tasks. \(X\) could range from 2 weeks to 2 months depending on the application and the scale of its dataset. These embeddings are then pushed to a feature store for online inference usage. Global block is the model deployed online for inference, which fetches all features from the feature store as well as the pre-populated meta embeddings to make the final prediction.

In terms of the cadence of refreshing both blocks, the model parameters \(\theta_{\text{meta}} \) and \(\theta_{\text{global}} \) are usually updated as per the regular retraining schedules of different applications. For most use cases, we do not frequently update them because we don't observe significant performance downgrade over time. However, we run the embedding generation flow daily, bootstrapping from \(\theta_{\text{meta}}\) to produce a fine tuned meta block \(\theta_{\text{meta}_i}\) using the latest user interaction data. The primary purpose of the meta block is to capture the fast evolution of user interest and provide personalization of the model. We expect the meta embeddings populated on a daily basis to provide a refreshed representation of each entity or user adapted from their latest engagement history. In our experiments on Push Notifications CTR prediction problem (pTap), updating the embeddings weekly instead of daily caused a drop in offline AUC by 0.5\%, indicating that refreshed embeddings are critical to maintaining meta learned model performance online.

\section{Offline Experiments}\label{sec:experiments}

In this section, we present offline results of LiMAML on various applications at LinkedIn. We begin by providing an overview of all applications and datasets in Section \ref{subsec:offlineoverview}. Then we use one of the applications as an example for a deep dive in Section \ref{subsec:ptapoffline}. We then summarize results of other applications in Section \ref{subsec:restoffline}. Some techniques to speed up LiMAML training is presented in Section \ref{subsec:speedup} and an ablation study is provided in Section \ref{subsec:ablation}.

\subsection{Applications and datasets}\label{subsec:offlineoverview}

We evaluate LiMAML across the following applications at LinkedIn.

\begin{itemize}
\item CTR Prediction of Push Notifications (\textbf{pTap}): This model predicts the probability of members tapping on a push notification card sent to their mobile devices from LinkedIn. We define the tasks at a user level, i.e. each recipient of the push notification becomes a new task.
\item CTR Prediction of InApp Notifications (\textbf{pClick}): This model predicts the probability of members clicking on a notification shown to the members on the in-app notifications tab within the LinkedIn app. For this model, we experiment with two different task definitions - a per-user task definition as well as a per-(user, notification type) task definition.
\item CTR Prediction of People You May Know Recommendations (\textbf{pInvite}): This model predicts the probability of a user sending a connection invite to the recommended person on My Network tab within the LinkedIn App. This model takes in two different entities and their features as input. The first entity is the user, also called as the inviter, who sends the connection invite and second entity is the invitee, who receives the connection invite. We experiment with both entities as the task definitions for this problem.
\end{itemize}

\begin{table}[htbp!]
\small
    \centering
    \begin{tabular}{ c|c|c } 
        \toprule
        \textbf{Application} & \textbf{Task Definition} & \textbf{Number of Tasks} \\ 
        \midrule
        pTap & per user & tens of millions\\
        pClick & per user & several millions\\ 
        pClick & per (user, notification type) & several millions\\
        pInvite & per inviter & tens of millions\\ 
        pInvite & per invitee & tens of millions\\
    \bottomrule
    \end{tabular}
    \caption{Application, Task Definition and Dataset Size across all applications.}
    \label{tab:offline_setup}
    \vspace{-2.0em}
\end{table}

We describe the scale of these datasets in Table~\ref{tab:offline_setup}. For all applications, the date range used to derive the training data is prior to the date range used for validation data. Similarly, the validation data is temporally before the test data. As each task has a varying number of samples, we also limit the number of samples per task to an upper bound by keeping only the most recent samples. When splitting the training dataset into support and query sets, we sort the samples for each task chronologically over time and assign the first 75\% of the samples to the support set, and the last 25\% to the query set. This is done to prevent any form of information leakage. Validation data is used for task-level fine-tuning and test data is used for final evaluation and metrics reporting. They can be regarded as the support and query set during inference time.

\subsection{LiMAML on pTap}\label{subsec:ptapoffline}

\subsubsection{Experiment Setup}
The baseline model for pTap uses an MLP network. We compare three different algorithms (described below) on this dataset. For each algorithm, we also wish to demonstrate the value of fine-tuning tasks on recent validation data.
\begin{itemize}[leftmargin=*]
\setlength\itemsep{0em}
\item \textbf{Vanilla Training:} In this approach, we train the neural network with regular gradient descent procedure using optimizers such as Adam. Usually, the samples in the training data are randomly shuffled and a mini-batch of samples are provided at a given time for training. However, in our experiments, we group the training data at a task-level first and then provide a mini-batch of tasks for training. This is done to provide a fairer and stronger baseline to compare with MAML. In the no-fine-tune scenario, we evaluate the same trained model on all tasks on the test data. In the fine-tune scenario, we use the trained model as bootstrap to take a few gradient descent steps on the validation data for each task, and evaluate this per-task model on the test data.
\item \textbf{Entire Network MAML:} In this approach, we use the MAML algorithm from \cite{finn2017} on the entire MLP network. In the no-fine-tune scenario, we evaluate the trained model output from the MAML algorithm on all tasks on the test data. In the fine-tune scenario, we use the trained model as bootstrap to do task adaptation on the validation data for each task, and evaluate this per-task model on the test data.
\item \textbf{LiMAML:} In this approach, we use the LiMAML algorithm described in \ref{alg:partial_maml_training} to meta learn part of the network. The meta block is set to a smaller MLP architecture in comparison to the global block. Referring to Figure ~\ref{fig:partial_maml_model}, meta features include all user-specific features, and, we use both meta features as well as the rest of the features as other features. The setting up of fine-tune and no-fine-tune are similar to Entire Network MAML, except that the task-adaptation (Line 4-8 of Algorithm ~\ref{alg:partial_maml_training}) is done only for the meta block on the validation data for each task.

\end{itemize}
For all experiments, we report AUC (Area under the ROC Curve) gain (or loss) in comparison to the contextual baseline on the test dataset. Across all our experiments, we usually observe the relative standard deviation of AUC gain over multiple runs to be less than 0.02\%. For the outer loop of both variants of MAML, we use the same optimizer as Vanilla Training baseline. In addition to the results on all tasks in the test data, we also slice the results into tasks with less than 25 samples and tasks with greater than 25 samples (in test data). This is done to highlight the efficacy of our approach even on tasks with smaller number of samples.

\subsubsection{Experiment Results}

We present the results of our experiments in Table~\ref{tab:pTapEntireOfflineResults}. As we can see, MAML with Fine Tuning gives significant gains in AUC in comparison to Vanilla Training without and with fine-tuning. This indicates that the model parameters output from the MAML algorithm provides a good bootstrap starting point to quickly and effectively adapt to any task. Additionally, the model parameters output from MAML are not expected to be at any optimal point, and hence we see that MAML without fine-tuning performs the worst among all approaches. Even though MAML gives the highest gains on tasks with larger number of samples, tasks with smaller number of samples do see significant gains as well. We also observe that LiMAML achieves comparable gains with Entire Network MAML, while additionally providing a production-friendly approach to personalization of large-scale online recommender models.

\begin{table*}[htbp!]
\small
\centering
\begin{tabularx}{\textwidth}{l|XX|XX|XX}
\toprule
\textbf{Algorithm} & \multicolumn{2}{l|}{\textbf{Vanilla Training}} & \multicolumn{2}{l|}{\textbf{Entire Network MAML}} & \multicolumn{2}{l}{\textbf{LiMAML }} \\
\midrule
\textbf{Fine Tune} & \textbf{No}  & \textbf{Yes} & \textbf{No} & \textbf{Yes} & \textbf{No} & \textbf{Yes}  \\
\midrule
All tasks in Test Data & baseline & +0.75\% & -0.18\% & +1.68\% & -0.43\% & +1.47\% \\
Tasks with Less Than 25 samples & baseline & +1.09\% & -0.29\% & +1.84\% & -0.36\% & +1.38\%\\
Tasks with More Than 25 samples & baseline & +1.63\% & -0.26\% & +2.83\% & -0.40\% & +1.64\%\\
\bottomrule
\end{tabularx}
\caption{AUC Gains (over baseline) From LiMAML Experiments on pTap.}
\label{tab:pTapEntireOfflineResults}
\vspace{-2.0em}
\end{table*}

\subsubsection{Comparison with Wide-and-Deep ID embedding models}
Here, we want to compare a MAML based personalization strategy with another popular personalization strategy in recommender models using ID embeddings. For this experiment, we construct a baseline pTap model with a wide-and-deep architecture \cite{cheng2016wide} with ID embeddings for every user ID. These embeddings are learnt via an embedding-lookup layer and are then concatenated with other dense features and passed through an MLP network. We compare two variants of LiMAML-based models - one with a learnable embedding-lookup layer as the meta block and the other with a small MLP as the meta block. The global block is kept identical to the baseline network without the ID embedding layer. In Table \ref{tab:pTapLPMResults}, we present the AUC results of the baseline network as well as the two LiMAML variants on test data. For the baseline network, we report results without fine-tuning since ID embeddings provide the personalization component. For the LiMAML variants, we report results on the test data after task adaption on validation data at a user-level. As we can observe from this experiment, LiMAML proves to be an effective strategy for personalizing a recommender model, with different meta block architectures - MLP layers and ID embeddings layers. In our future work, we plan to explore using more complex architectures, such as transformers \cite{vaswani2017attention}, as the meta block, to understand the performance across different architectures.


\begin{table}[htbp!]
\small
\centering
\begin{tabular}{c | c | c}
\toprule
\textbf{Global Block} & \textbf{Meta Block} &   \textbf{Test AUC Gain}  \\
\midrule
MLP + ID Embeddings & N.A. & baseline \\
MLP & ID Embeddings & +1.30\% \\
MLP & MLP & +1.36\% \\
\bottomrule
\end{tabular}
\caption{LiMAML compared with ID embedding based approach.}
\label{tab:pTapLPMResults}
\vspace{-2.0em}
\end{table}

\subsection{LiMAML on other applications}\label{subsec:restoffline}
In Table~\ref{tab:rest_offline}, we list the performance of LiMAML with different task definitions across several other applications. Baseline model for all experiments use the deep neural network architecture for that application. We train the baseline model using Vanilla Training by providing a mini-batch of tasks (samples grouped at a task level). We observe consistent AUC gains from LiMAML (with task adaption) for most applications in comparison to the baseline (without fine-tune). This indicates that the proposed approach provides a robust personalization paradigm for all recommender system models. Additionally, we also learn the significance of choosing the right task definition in meta learning. As seen in pClick, a broader task definition at a user level provides little to no gains, whereas a fine-grained task definition at a (user, notification type) level has a higher gain. We have observed such patterns across many applications.

\begin{table}[htbp!]
\small
    \centering
    \begin{tabular}{ c|p{3cm}|c } 
        \toprule
        \textbf{Application} & \textbf{Task Definition} & \textbf{Test AUC Gain} \\ 
        \midrule
        pInvite & per inviter & +10.56\%  \\ 
        pInvite & per invitee & +0.58\% \\ 
        pClick & per user & +0.01\%  \\ 
        pClick & per (user, notification type) & +0.47\%  \\ 
    \bottomrule
    \end{tabular}
    \caption{AUC Gains (over baseline) from LiMAML Experiments on all other applications.}
    \label{tab:rest_offline}
    \vspace{-2.0em}
\end{table}

\subsection{Training speed-up} \label{subsec:speedup}
LiMAML training is computationally more expensive compared to Vanilla Training due to the additional inner loop task adaptation step resulting in Hessian-vector product computation ~\cite{finn2017}. On pTap, training with 5 inner loop iterations resulted in 221\% increase in training time over Vanilla Training (see Table ~\ref{table:task-iterations} in Section~\ref{sec:inner-loop-iterations}). We present some techniques which helped us significantly reduce the training time. 

Firstly, we increased the number of tasks per batch per GPU from 32 to 128. As evidenced in \cite{goyal2017accurate}, simply increasing the batch size can potentially introduce training instability. Therefore, we employed following techniques to mitigate the challenges with large batch training:
\begin{itemize}[leftmargin=*]
\item We linearly scaled the global learning rate as we scaled the number of tasks per batch \cite{goyal2017accurate}. We experimented with different values around the scaled learning rate to choose the optimal setting.  
\item We clipped the gradient norm to $1.0$ during the outer loop. 
\item We utilized global learning rate scheduling with warm-up and decay.  
\end{itemize}
With these changes, LiMAML could achieve the same offline metrics with 46.85\% reduction in training time. We observed that learning rate scheduling and gradient clipping play a pivotal role in achieving faster convergence speed. 

Secondly, we utilized multi-GPU training in a data parallel paradigm with gradient synchronization between different GPUs. The gradient synchronization is done only during the outer loop but not during the inner loop updates. Multi-GPU training along with the large batch size optimization gave us a 64.28\% reduction in training time. 

Thirdly, we realized that the number of inner loop gradient updates, \(n\), significantly affects the training time. We didn’t observe a significant difference in AUC gain as we vary \(n\). Hence, we change \(n\) to 1 to achieve further speed-up in training (refer to Appendix \ref{sec:inner-loop-iterations} for details). Overall we achieve 77\% training speedup by applying all the above techniques. We summarize these results in table \ref{table:training-time}.

\begin{table}[htbp!]
\small
    \centering
    \begin{tabular}{p{4.5cm}|c}
        \toprule
        \textbf{Technique} & \textbf{Train Time Reduction} \\ 
        \midrule
        LiMAML & baseline \\
        \hspace{.5em}+Large batch size, learning rate  \hangindent=1em 
        optimization, gradient clipping  & -46.85\%  \\
        \hspace{1.5em}+Multi-GPU training & -64.28\%  \\
        \hspace{2em}+Single inner loop iteration & -77.19\%  \\
    \bottomrule
    \end{tabular}
    \caption{LiMAML training time reduction after applying various techniques.}
    \label{table:training-time}
    \vspace{-2.0em}
\end{table}

\subsection{Ablation study}\label{subsec:ablation}
\label{ablation-study}
In this section, we provide some insights on the sensitivity of different hyper-parameters and some observations from our experiments with LiMAML.
\begin{itemize}[leftmargin=*]
\item Among the hyper-parameter choices, we observed a relatively higher sensitivity to global learning rate, which requires some tuning to see gains. In Appendix ~\ref{sec:hyperparam-tune}, we present some experiments on pTap and pClick to illustrate this.
\item We experimented with applying dropout after each MLP layer. However, even a dropout rate as low as 0.1 resulted in a drop in test AUC. We postulate that this is the result of using large training dataset with tens or hundreds of millions of tasks which prevents the meta learn model from overfitting. More details about this experiment is presented in Appendix ~\ref{sec:dropout}.
\item In Algorithm \ref{alg:partial_maml_inference}, we score the latest sample for each task with the fine-tuned meta block to produce meta embeddings. We also investigated the impact of different aggregation methods such as max, mean pooling to produce these embeddings and they all performed poorer than using the latest sample. More details are presented in ~\ref{sec:pooling}.

\end{itemize}

\section{Online Experiments}\label{sec:online}

\subsection{Online A/B test results}
We have deployed LiMAML based pClick and pTap models for evaluating the propensity of clicking and tapping on notifications. These predicted CTRs are important components in the offline reinforcement learning based decision making system \cite{prabhakar2022multiobjective} to make notification send/drop decisions. Both pClick and pTap models were experimented separately online with their respective deep neural network as the baselines. We evaluated the performance of these models on the following metrics.
\begin{itemize}[leftmargin=*]
    \item \textbf{Weekly Active Users (WAU)}: The number of unique members who have visited the LinkedIn site within a seven-day period. This is one of the most important metrics to drive long term value on our platform.
    \item \textbf{Notifications Click-Through Rate (CTR)}: The average click-through rate of notifications sent to the members on a daily basis. This is an important metric to measure the relevance of notifications sent to the members. 
\end{itemize}

\begin{table}[htbp!]
\small
\centering
\begin{tabular}{c|c|c}
\toprule
 & \textbf{CTR} & \textbf{WAU} \\
\midrule
pClick & +2.0\% & +0.1\% \\
pTap & +0.6\% & +0.2\% \\
\bottomrule
\end{tabular}
\caption{LiMAML relative metrics improvement from online experiments.}
\label{tab:pTapOnlineResults}
\vspace{-2.0em}
\end{table}

All A/B tests have been conducted online with production traffic, lasting a week. The results are reported in table ~\ref{tab:pTapOnlineResults}. All numbers are statistically significant $(p\text{-value} < 0.05)$. As shown in the table, LiMAML based models have not only improved the relevance of notifications sent to the user, but also driven long term value to members on LinkedIn.

\subsection{Analysis of online results}\label{sec:online_analysis}
Frequent members of the platform are usually over-represented in the training datasets across most applications. Hence, we have seen modeling efforts seldom showing improved performance on members with little to no data in the training set. However, with a meta learning based approach, we expect to see personalization of models even on tasks with very few (>=1) samples. We evaluate online model performance on three select user cohorts who are infrequent members of LinkedIn, whom we have seen in our data to have very few samples.
\begin{itemize}[leftmargin=*]
    \item \textbf{Cohort 1}: Members who have visited at least once a week in one of the past four weeks. These users can be roughly regarded as Monthly Active Users.
    \item \textbf{Cohort 2}: Members who have not visited in the past 28 days but have visited in the last 3 months.
    \item \textbf{Cohort 3}: Members who have signed up newly on our platform in the last four weeks. These members' data have not been included in training the LiMAML model.
\end{itemize}

\begin{table}[htbp!]
\small
\centering
\begin{tabular}{c|c|c|c|c}
\toprule
& \textbf{Metric} & \textbf{Cohort 1} & \textbf{Cohort 2} & \textbf{Cohort 3} \\
\midrule
\multirow{2}{*}{pClick} & CTR & +7.7\% & +4.9\% & neutral \\
 & WAU & +0.4\% & neutral & neutral\\ 
\hline
\multirow{2}{*}{pTap} & CTR & +3.7\% & +11.2\% & +5.3\% \\
 & WAU & +0.5\% & +0.7\% & neutral\\ 

\bottomrule
\end{tabular}
\caption{For cohorts with fewer data points, meta learning shows significant relative metrics lift.}
\label{tab:IFResults}
\vspace{-2.5em}
\end{table}

In Table~\ref{tab:IFResults}, we present online A/B test results for these three selected user cohorts. From the table, we can observe that LiMAML models have shown huge online CTR gains as well as long term engagement (WAU) on these cohorts. 
These gains are large in comparison to what we usually observe for these cohorts in our modeling experiments. Additionally, in one case, we see significant gains in Cohort 3, whose tasks are not present during training LiMAML. This is stemming from the fact that meta learning significantly uplifts the model performance even for members with little to no data, thereby enhancing user segments which were previously under-optimized by the global models. These results also indicate that unlike ID embedding based approaches which usually require a decent amount of data per entity, meta learning can adapt very well to any new or existing task given a few data points, making it a more generalized and effective personalization strategy.

\section{Conclusion}\label{sec:conclusion}
In this paper, we introduced a meta learning method called LiMAML, which provides a generalized framework for personalization of recommender models. We have done extensive experiments to demonstrate significant offline and online metric lifts against state-of-the-art recommendation models when the framework is deployed to different LinkedIn applications. We have also shown the efficacy of our approach on tasks with very little data.

For future work, we want to explore different extensions of this approach, along with new applications. Firstly, we want to integrate LiMAML with our in-house foundation models, such as Large Language Models, Reinforcement Learning Agents and Graph Neural Networks. Secondly, many applications contain a flavor of multiple entities for which we need to achieve personalization simultaneously. For example, an ad CTR prediction problem might want a model personalized per user as well as per advertiser, but a task definition of user-advertiser pair may not make sense. Similarly, for a notifications CTR prediction problem, we might want a user task as well as a user-notification type task simultaneously. We are exploring ways to extend LiMAML for multiple simultaneous task definitions, either via multiple meta-blocks or via intelligently combining different task distributions. Thirdly, we are exploring different architectures such as a transformer for the meta block. Given that the meta block is trained on a chronological sequence of user interaction data, a sequential architecture might give higher gains while personalizing on such a data.

\section{Acknowledgments}\label{sec:conclusion}
The authors would like to thank Mohsen Jamali, Shipeng Yu, Angus Qiu, Viral Gupta, Akashnil Dutta, Parag Agrawal, Xiaobing Xue and others who collaborated with us, and Kenneth Tay, Ruoying Wang, Ankan Saha for reviewing the paper and providing insightful suggestions.

\bibliographystyle{ACM-Reference-Format}
\bibliography{bibliography}

\appendix
\section{INFORMATION FOR REPRODUCIBILITY}\label{sec:reproducability}

\subsection{Results for different inner loop iterations}
\label{sec:inner-loop-iterations}
Table \ref{table:task-iterations} illustrates the increase in training time for pTap application when we introduce meta learning over vanilla training. When increasing the inner loop iterations \(n\), the training time increases roughly linearly while AUC gains remain the same. We have similar observation on pClick applications.

\begin{table}[htbp!]
\small
\centering
\begin{tabular}{>{\centering\arraybackslash}p{1.6cm} | c | c}
\toprule
\textbf{Inner Loop Iterations} & \textbf{Train Time Increase } &   \textbf{Test AUC Gain}  \\
\midrule
Baseline & baseline & baseline \\
1 & +104.98\% & +1.47\% \\
2 & +130.35\% & +1.50\% \\
3 & +162.40\% & +1.50\% \\
4 & +185.77\% & +1.50\% \\
5 & +221.16\% & +1.51\% \\
\bottomrule
\end{tabular}
\caption{LiMAML training time increase and Test AUC gains for different values of inner loop updates on pTap.}
\label{table:task-iterations}
\end{table}

\subsection{Results with different dropout rates}
\label{sec:dropout}
In this experiment on pTap, we apply dropout to every layer of the neural network. With dropout, we observe a drop in Test AUC gains. As presented in Table ~\ref{table:dropout}, test AUC gains drop from +1.47\% to +1.44\% with a minimal dropout rate of 0.1. As the dropout rate is increased, we observed the drop in AUC consistently increasing.

\begin{table}[htbp!]
\small
\centering
\begin{tabular}{c | c}
\toprule
\textbf{Dropout rate} & \textbf{Test AUC Gain}  \\
\midrule
Vanilla training & Baseline \\
0. & +1.47\% \\
0.1 & +1.44\% \\
0.2 & +1.40\% \\
0.4 & +1.26\% \\
0.6 & +0.88\% \\
\bottomrule
\end{tabular}
\caption{pTap LiMAML Test AUC gains for different dropout rates with respect to vanilla training baseline.}
\label{table:dropout}
\vspace{-2.0em}
\end{table}

\subsection{Hyperparameter tuning experiments}
\label{sec:hyperparam-tune}
On the pTap LiMAML model, we fix the hyperparameter configuration and experiment with different task learning rate values. As evident from Table ~\ref{table:ptap-task-lr-tune}, tweaking task learning rate (0.005 to 10) around the best model configuration has insignificant drop on the test AUC gains. We have similar observation for pClick LiMAML (user, notification type) model as shown in Table ~\ref{table:pclick-hyperparam-tune}. 
We perform similar experiment over global learning rate values to observe the effects on test AUC gains. The resulting test AUC gains are shown in Table ~\ref{table:ptap-hyperparam-tune}.

We observe that in comparison, the task learning rate requires less tuning and is usually set to 100x or 1000x orders of magnitude higher than global learning rate. This observation is consistent with high task learning rates values used in other MAML works ~\cite{finn2017, raghu2019rapid}.  

\begin{table}[htbp!]
\small
\centering
\begin{tabular}{c | c}
\toprule
\textbf{Task Learning Rate} &   \textbf{Test AUC Gain}  \\
\midrule
Vanilla training & Baseline \\
0.001 &	0.44\% \\
0.005	& 0.46\% \\
0.01	& 0.41\% \\
\textbf{0.05}	& \textbf{0.47\%} \\
0.1	& 0.45\% \\
0.5	& 0.45\% \\
1	& 0.44\% \\
5	& 0.45\% \\
10	& 0.46\% \\
20	& 0.45\% \\
50	& 0.42\% \\
\bottomrule
\end{tabular}
\caption{pClick (user, notification type) LiMAML task learning rate tuning. Global learning rate is set to .0006 for the above experiments. Test AUC gain is over vanilla training baseline.}
\label{table:pclick-hyperparam-tune}
\end{table}

\begin{table}[htbp!]
\small
\centering
\begin{tabular}{c | c}
\toprule
\textbf{Task Learning Rate} &   \textbf{Test AUC Gain}  \\
\midrule
Vanilla training & Baseline \\
0.005	&	1.44\% \\
0.01	&	1.45\% \\
0.05	&	1.47\% \\
\textbf{0.1} & \textbf{1.47\%} \\
0.5	&	1.45\% \\
1	&	1.45\% \\
10	&	1.46\% \\
\bottomrule
\end{tabular}
\caption{pTap LiMAML task learning rate tuning. Global learning rate is set to .0012 for the above experiments. Test AUC gain is over vanilla training baseline.}
\label{table:ptap-task-lr-tune}
\end{table}

\begin{table}[htbp!]
\small
\centering
\begin{tabular}{c |  c}
\toprule
\textbf{Global Learning Rate} & \textbf{Test AUC Gain}  \\
\midrule
Vanilla training & Baseline \\
0.00095 & +1.47\% \\
0.0009	& +1.47\% \\
\textbf{0.0012}	& \textbf{+1.47\%} \\
0.0015	& +1.46\% \\
0.003	& +1.43\% \\
0.004	& +1.41\% \\
0.005	& +1.37\% \\
0.006	& +1.40\% \\

\bottomrule
\end{tabular}
\caption{pTap LiMAML global learning rate tuning. Tasks learning rate is set to 0.1 for these experiments. Test AUC gain is over vanilla training baseline.}
\label{table:ptap-hyperparam-tune}
\end{table}

\subsection{Aggregation strategies}\label{sec:pooling}
In our experiments, we score the latest sample for each task with the fine-tuned meta block to produce meta embeddings. Alternatively, we also tried the following strategies (Max, Mean, Cos) to produce these embeddings. : 
    \begin{itemize}
    \item \textbf{Max:} Max pooling across all samples for each task.
    \item \textbf{Mean:} Mean pooling across all samples for each task.
    \item \textbf{Cos:} Cosine similarity weighted mean pooling, where we take a weighted mean across all the samples for each task with the weights set to the cosine similarity between that sample and the latest sample for the task.
    \end{itemize}

As we can see from Table \ref{tab:pooling results}, all these aggregation strategies performed poorer in comparison to deriving the meta embeddings from the latest sample for each task.

\begin{table}[htbp!]
\small
\centering
\begin{tabular}{c|c|c|c}
\toprule
 & \textbf{Max} & \textbf{Mean} & \textbf{Cos} \\
\midrule
AUC & -0.06\% & -0.12\% & -0.10\% \\
\bottomrule
\end{tabular}
\caption{AUC Gains using different pooling methods for Meta Embedding Generation.}
\label{tab:pooling results}
\vspace{-2.0em}
\end{table}

\end{document}